\documentclass[11pt]{article}
\usepackage[utf8]{inputenc}
\usepackage{marvosym} 
\usepackage{amssymb,latexsym,amsmath,color}
\usepackage[colorlinks,linkcolor=blue,urlcolor=blue,citecolor=black,plainpages=false,pdfpagelabels,breaklinks]{hyperref}
\usepackage[pdftex]{graphicx}
\usepackage{physics}
\usepackage{ulem}

\renewenvironment{enumerate}{\begin{list}{}{\rm \labelwidth 0mm
\leftmargin 5mm}} {\end{list}} 

\newcommand{\ita}{\textit}

\newcommand{\lra}{\leftrightarrow}

\title{Identical particles in quantum mechanics: favouring the Received View}
\author{{Décio Krause}\thanks{Department of Philosophy, Federal University of Santa Catarina and Post-Graduate Program in Logic and Metaphysics, Federal University of Rio de Janeiro. Partially supported by CNPq.} \and {Jonas R. B. Arenhart}\thanks{Institut Wiener Kreis, Vienna. Leave from the Department of Philosophy, Federal University of Santa Catarina. Partially supported by CNPq.}}
\date{April 2020}

\begin{document}
\maketitle

\begin{abstract}
 The so-called Received View (RV) on quantum non-in\-di\-vidu\-al\-ity states, basically, that quantum particles are not individuals. It has received an amount of criticism in the recent literature, most of it concerning the relation between the RV and the relation of identity. In this chapter we carefully characterise a family of concepts involved in clarifying the view, indicating how the very idea of failure of identity, commonly used to define the RV, may be understood. By doing so, we hope to dissipate some misunderstandings about the RV, which shall also be seen as evidence of its tenability. 

\medskip 
\noindent 
Keywords: non-individuals; Received View; identity; quantum mechanics.
\end{abstract}

\bigskip
\hfill{
\begin{minipage}{8cm}
{\small ``I am the same electron that spoke to you before."

``You cannot be!" exclaimed Alice. 

``I saw that electron go off in a different direction. Perhaps he was not the same one I was talking to before?"

``Certainly he was."

``Then you cannot be the same one," said Alice reasonably. 

``You cannot both be the same one you know."

``Oh yes we can!" replied the electron. 

``He is the same. I am the same. We are all the same, you know, exactly the same!"

``That is ridiculous," argued Alice. 

``You are here beside me, while he has run off somewhere over there, so you cannot both be the same person. One of you must be different."

``Not at all," cried the electron, jumping up and down even faster in its excitement.

``We are all identical; there is no way whatsoever that you can tell us apart, so you see that he must be the same and I am the same too." \\ Robert Gilmore, \ita{Alice in the Quantumland}} \cite[p.30]{gil95}
\end{minipage}
}

\section{Introduction}

The Received View (RV) on quantum particles' non-individuality says that quantum objects are \ita{non-individuals} of some sort (see \cite{frekra06} and the references therein; see also \cite{are17} for further discussion). As an attempt to characterise the view, it does not say anything very specific, given that all it is doing is to deny that quantum particles are \ita{individuals}. It provides, however, a kind of recipe for versions of the RV, which get their meaning depending on how one frames the details concerning individuality and lack thereof. In the absence of a precise understanding of what individuality means, and how something may fail to be an individual, this leaves a lot of space for misunderstanding, and the aim of this paper is to explain how much of the potential criticism to the RV may be properly dealt with. 

Addressing such criticism will require that we clearly characterise the RV. Instead of defending one particular approach to the RV (it may be characterised in distinct terms, see \cite{are17}), we shall approach the core concepts involved in articulating the view, and show that a proper understanding of them will work perfectly well to dispel some of the doubts that are constantly surrounding it. This shall also be a nice opportunity to discuss some methodological points on the metaphysics of quantum mechanics related to the RV, and which have only appeared thanks to discussions on the RV. We shall suggest, in the end, that quantum mechanics indicates non-individuality as a natural company for the theory, despite the current criticism to the view.

This paper is organised as follows. In the next section, we make an overview of core metaphysical and logical concepts involved in articulating the RV. These concepts are those that are typically called forth also in criticisms of the RV, so, a proper understanding of them is important if one is to address the RV properly. In section \ref{sec3} we attempt to bring to light some of such confusions, and how the proper treatment of the concepts may avoid them, or suggest alternative roads to dealing with these concepts, mainly when issues concerning isolation of quanta are involved. Next, in section \ref{sec4}, we describe CTI, the classical theory of identity in order to see what it entails and why it brings problems for the RV, given its most common formulation related to the denial of identity for some entities. In section \ref{sec5} we discuss the famous analogy of the money in a bank account, which has both been used to advance the RV, and to criticise it in more recent literature. Finally, in section \ref{sec6} we present the main outlines of quasi-set theory, a theory that codifies some of the basic claims of the RV, and grounds our claim that identity may be associated with individuality, and may be dispensed with for some entities (quantum non-individuals, of course!). A brief comment on multisets and on fuzzy sets is given for comparison. We conclude in section \ref{sec7}, with a small coda on metaphysical theory choice in the specific case of quantum mechanics and theories of individuality.

\section{Some of the background concepts to the RV}\label{sec2}

From a metaphysical point of view, the first notion that one must grasp in order to make sense of the RV is the notion of an individual. Non-individuals, in the sense advanced by the RV, are neither individuals nor universals or non-particular items. Rather, they are traditionally thought of as \ita{particulars lacking a principle of individuality}. Obviously, this does not mean that the RV suggests that everything is not an individual: the claim, again, is that quantum particles are not individuals.\footnote{Certainly, one could claim that other kinds of entities, such as clouds, may not be individuals. However, in this paper we shall confine ourselves to non-individuals in quantum mechanics, given that one may see the theory itself as providing us reasons to believe that such entities lack individuality.} 

`Individual' is a technical term in metaphysics (see Lowe \cite{low03} for a discussion of the standard approaches to individuality). Usually, in saying that something is an individual, philosophers use to point to some \ita{principle of individuality} and say that, since something obeys such a principle, that is, once this something `has individuality', it is an individual. As we shall see, there are further things to consider.\footnote{More specifically, we shall point to the necessity of distinguishing among three usually fundamental notions, namely, identity, individuality, and individuation.} The fact is that there is not a clear consensus on what `individual' means exactly, given that distinct rival proposals have been advanced on what should count as an individuality principle, and the debate on the metaphysical principle of individuality is still alive (as other metaphysical debates). The target notion that one tries to characterise by a principle of individuality is the idea of \ita{one} unity of a given kind (a chair, a car, a person, a horse, etc.), with the principle singling out that particular individual as distinct from any other individual of the same kind. An individual, then, is a particular item satisfying such a principle, and so that it may have properties and can enter in relations with other individuals. Given that a principle of individuality also accounts for the differences between individuals, they can be \ita{counted} in the traditional sense, meaning that given a finite collection of them,\footnote{In this discussion, we consider only finite collections.} we may apply the classical definition of counting, which requires that we define a one-to-one function from the set they form with a unique finite ordinal (in von Neumann's sense). They may bear names and, importantly, these names act as rigid designators, meaning that the individuals can (in principle) be referred to (for instance, by their unambiguous names) as such in different contexts. In this sense, the metaphysical concept of individuality requires, for its formulation and application, a close connection with the concept of identity (see also Bueno \cite{bue14}).

This general notion of individuality has been developed in distinct directions, giving rise to distinct theories of individuality. In broad terms, two rival approaches are \ita{the Leibnizian qualitative} approach to individuality, grounding the individuality in terms of the qualities the individual has, and the so-called \ita{Transcendental Individuality} approach, which grounds the individuality in terms of some non-qualitative ingredient (see French and Krause \cite[chap.1]{frekra06}; see also Benovsky \cite{ben16} for doubts on the rivalry between the views). The Leibnizian approach frames individuality exclusively in terms of the qualitative properties instantiated by a particular; basically, one individual's individuality is completely characterised by the fact that each individual is unique on what concerns the collection of qualities it possesses or instantiates. The uniqueness in case is achieved by endorsement of the famous Leibniz's \ita{Principle of the Identity of the Indiscernibles} (PII), which states that there are no \ita{two} absolutely indiscernible things: the agreement in all properties requires that there is numerical uniqueness, and so complete characterisation by collection of properties counts as a principle of individuality. 

Of course, one could doubt the underlying idea that a qualitative description of an individual can be uniquely attributed to a single individual, so that it would be possible for two \ita{individuals} to share every property, failing the PII. In this case, qualitative identity would not be enough for numerical identity, and the principle of individuality would have to be provided by something transcending the qualities of the individuals. This approach requires that there exists something beyond the properties of an individual that makes it the individual it is, and so the principle of individuality would be based on the existence of some form of substratum (a particular ingredient added to the characterisation of the individual and which confers it its individuality), a haecceity (a non-qualitative property instantiated solely by the individual it characterises), or whatever metaphysical posit that can work for those purposes; philosophers, following Heinz Post, call this strategy the `transcendental individuality' approach (\cite{redtel91}, \cite[chap.1]{frekra06}). 

Classical physics is compatible with both approaches. It assumes Max\-well-\-Boltzmann statistics, where permutations of qualitatively indiscernible things (like two electrons) lead to \ita{different} states. So, although they can be in principle indiscernible by their qualities, the particles have something granting them individuality, and it is suggested that it is their individuality which operates on the background to account for the difference of the states before and after the permutation (given that no quality of the particles could do that in such cases). A Leibnizian may account for that by appealing to the \ita{Impenetrability Principle}, which grants a unique spatial location to any particle; in this case, the notion of quality allowed by the Leibnizian must be extended to allow for spatial locations to be seen as a quality and act as a principle of individuality. Transcendental Individuality approaches needs no such appeal to spatial location as a special quality, given that they already posit something transcending every quality and relation an object may have. In this sense, the metaphysics of individuality in classical mechanics gets underdetermined by classical physics (see French and Krause \cite[chap.2]{frekra06}).

That accounts for the metaphysical concept of individuality. Notice that by advancing a principle of individuality, one has answered the metaphysical question of individuality: what is it that makes an individual precisely the individual it is? Qualitative and non-qualitative approaches advance specific answers to that. A major difference in these approaches concerns how they allow one to address a different concern about entity \ita{identification}. Qualitative approaches to individuality, by framing individuality in terms of qualities, allow that, at least in principle, we uniquely identify an individual as the one having such and such properties. Given that these approaches assume that an individual is uniquely characterised by its qualities, this can be done, again, at least in principle. Non-qualitative principles cannot do that, \ita{i.e.} ground identification claims on individuality, given their assumption that distinct individuals may be qualitatively identical. This epistemic act of singling out an entity as the object of our sensory attention or reference is what we shall call \ita{individuation}. 

Individuality and identification have a close relationship with another important concept, which also features in discussions of these issues in the literature and on the RV: \ita{identity}. What it is meant by `identity'? This is a perennial philosophical question. Philosophers of physics, as philosophers in general (in particular the critics of the RV), assume some informal (and quite vague) notion of \ita{numerical} identity: individuals $a$ and $b$ are identical (we write $a = b$) if and only if there are no two individuals at all, but just one, which can be referred to indifferently by either $a$ or $b$. Systems of logic attempt to bring systematic clarity to the relation of identity, although controversy appears here too (for instance, in the resulting distinctions between first-order versus second-order versions in classical logic, and necessary or contingent identity in modal logic, to mention just two). The relation is important for us because it has been frequently used to characterise \ita{non-individuals}: it is typically advanced that a non-individual is an entity for which identity does not apply. That is, non-individuals are entities for which identity, in the sense of numerical identity, does not hold. But what is the relation between such failure of identity and failure of individuality?  

According to some forerunners of quantum physics, quantum particles cannot be seen as individuals, understood as entities with well defined identity conditions (the story can be seen in \cite[chap.3]{frekra06}; but see also \cite{arebuekra19}). Then, they would be \ita{non-individuals} in this sense, of lacking anything that makes them precisely the individual they are, there is no such ingredient available for a quantum particle. This was taken seriously by the most well-developed version of the RV. The most well-known approaches to the RV (as advanced in \cite{frekra06}) intends to understand and to analyse formally the lack of individuality as a failure of the notion of identity when applied to certain entities. According to the RV, then, the very notion of identity would not be applicable to all objects. Formally, expressions like $a = b$, for $a$ and $b$ terms designating specific objects (quantum entities), would not be well-formed, and so the language would be unable to express their identity (or difference). But, as the reader may be wondering, even so they could have identity, since this could be just a lack of expressive capacity of the language, so being an epistemological limitation. As we shall show below, this can be accepted as being the case, \ita{if} we use classical logic in the metamathematics, a fact that most critics of the RV don't take notice of.\footnote{Really, they always presuppose classical logic and a standard set theory, encompassing the classical theory of identity (CTI) to be seen soon.} It was precisely this question that motivated the development of an alternative framework for semantic considerations, so that $a = b$ (or $a \ne b$) could not be proven even in the metamathematics, namely, the so-called \ita{quasi-set theory}, to be mentioned below (see also Arenhart \cite{are14}).

The very idea that a formal background is always called forth is a must in these discussions. It puts the desirable order on a metaphysically important concept, and allows us to make a better sense of what it would mean for something to lack such concepts, or for such concepts to fail to apply. This can be understood in diverse ways, and the proper formalisation helps us keeping track of what is being advanced and what is not allowed. The first thing to be acknowledged is that this metaphysical notion of identity cannot be properly captured by a first-order formal system \cite{hod83} (see also \cite{kraare18}, specially p.287). Even so, we still must count with \ita{some} theory of identity, and usually we choose the theory ascribed by classical logic and standard mathematics, as we shall see below. 


\paragraph{Remark} Of course, identity and its relation to individuality demands a further discussion. So far, we have said that traditionally, non-individuality has been framed as a kind of lack of identity. That brings with it further questions concerning identity that shall not concern us here. We can say that Dr. Jekyll has an identity when he is Dr. Jekyll, but loses his identity when he becomes Mr. Hyde and vice-versa. In psychology, we know related cases of people who have different personalities in different times. All of this is relevant for the discussion of identity, but we are not discussing identity in time (transtemporal identity). Anyway, Dr. Jekyll and Mr. Hide are \ita{the same person}, although assuming different \ita{personalities}. We assume that individuals present \ita{genidentity}, as in standard physics, in the sense of being able to be \ita{re-identifyed} as such in other contexts. 

Furthermore, non-individuals obey permutation laws, such as the Indistinguishability Postulate (see \cite{redtel91}, French and Krause \cite[chap.4]{frekra06}), which roughly says that a permutation of non-individuals (better: permutations of non-indivividuals' labels) does not conduce to a different state of affairs. They cannot be counted in the standard sense, put in some order in the standard sense, for they do not have identity. We end up not being able to define a one-to-one function from a set of such entities to a finite ordinal, for to which non-individual we should ascribe the number 2 if they cannot be identified? Anyway, their collections may have a cardinal, for in many cases we refer to `six electrons so and so', `two quarks up', and so on. These features of a non-individual must be taken into account into any version of the RV.

\section{Confusions involving individuality and individuation}\label{sec3}

The RV is accused of not reflecting present day physics, since in some situations particles can be isolated, hence, according to some critics, presenting identity and individuality, while the RV says that such entities do not present identity conditions. This is a conclusion that mixes the different concepts of \ita{identity}, \ita{individuality}, and \ita{individuation}.\footnote{These notions were already differentiated in \cite{kraare18}.} So, before we address directly the problems that identity may be seen as raising in connection to the RV, and some of the criticism associated with it, it will be useful to consider some of the confusions that a conflation of individuality and individuation have caused to the RV. Our claim is that once these concepts are properly understood, as we suggested in the previous section, no difficulty appears for the RV (see also \cite{kraare18}).

Non-individuals lack individuality, but they need not lack individuation conditions (recall the difference, as stated in the previous section). That is, some non-individuals may be identified, in the epistemic sense of individuation, and still not derive any principle of individuality in the metaphysical sense from such an identification. The fact that a positron, for instance, is trapped in the laboratory does not entail that we have a well defined principle of individuality in our metaphysical pantheon. The situation is perfectly the same if the particle is taken to be a non-individual.

Part of the criticisms to the RV come, we guess, from a misunderstanding of which of these characteristics a non-individual lacks. Non-individuals can certainly be isolated (as a positron in a trap), taken as one of a kind (one electron), bear properties (electrons have a certain mass, a specific electric charge, and so on), may enter in relations with other non-individuals (to have spin opposite to) and with individuals, but still fail to have other characteristics. For instance, they cannot always be re-identified as such from context to context. An electron released from an atom by a process of ionisation cannot be ever recognised again; one cannot hope to find an electron at a later time and say: `that is exactly the electron that was released before'. So, a given name (as `Priscilla') does not act as a rigid designator, and cannot be used to identify the positron as \ita{that} positron that was trapped one day (see below). 

Individuation is an epistemological notion, and it is closely related to the notion of \ita{isolation}; we isolate an entity for some purposes. Going directly to our case study, we may say that a certain quantum object (a positron called Priscilla \cite{deh89}, a barium atom called Astrid \cite{deh89}, a strontium atom \cite{zac18}, etc.) is \ita{individuated}, or \ita{isolated} by some trapping device.\footnote{Hans Dehmelt won the Nobel Prize of 1989 for trapping quanta, and the same happened with Serge Haroche and David Wineland in 2012. But, as we have argued in \cite{kra11}, these experiments neither confer to the trapped quanta a status of individuals, nor necessarily confer them identity. See below. Notice that we have avoided to use the article `the', preferring to speak of \ita{a} positron, \ita{a} barium atom, \ita{a} strontium atom.}  Of course the asymmetries of the laboratory provide the means for us to provide individuation for the trapped quantum, but not necessarily an individuality, in the sense that it also confers the metaphysical explanation of what the entity is. It is simply isolated. But some physicists say that due to this situation, the trapped entity acquires an identity card, or something like individuality. Does it?

From our previous discussion on individuality, the conclusion just doesn't follow. Isolating an entity does not confer to it individuality in the sense that we would be able to say that due to the isolation, we have a principle of individuality, not even in qualitative approaches to individuality. Basically, to be trapped in a quantum trap, if quantum entities are individuals, the principle of individuality must be already operating (the individual must exist as an individual before it gets trapped). The trap cannot operate as a temporary individuating principle, in any metaphysically robust sense. There is no such thing. The particle does not become the particle it is, different from any other, just because it is inside a trap. In other words: the fact of being trapped does not answer the concerns a principle of individuality should answer. Even more: trapping is just temporary, so that, if it could confer individuality to a particle, this individuality would be temporary too. That doesn't make sense for individuals. Julius Caesar was not Julius Caesar in the working days and Pompey in the weekends. While existing, he was Julius Caesar all the time, even if sometimes used a fantasy to avoid being recognised. But even in this case he was Julius Caesar, and if he fought against Vercingetorix, it was Julius Caesar who did it, and not Pompey. 

From the perspective of the distinction presented in the previous section, quantum objects, when trapped, instantiate what Toraldo di Francia called a \ita{mock individuality} (we could add that they also have a \ita{mock identity}), a term Dennis Dieks considers `disparaging' \cite{die20}. This believe is due, we guess, to the (again) confusion among identity, individuality, and individuation (isolation). The trapped quantum object is isolated by the trap, and there is a condition of \ita{individuation} operating on the background, namely, the possibility for us to say that \ita{that} object in the trap is a quantum object of a specific kind, a positron say. But it can (by hypothesis) leave the trap, and get mixed with other objects of the same kind. In this case, even the mock individuality is lost. That means that the trap does not confer individuality to the trapped particle, but only a temporary individuation condition provided by the condition of the laboratory apparatus. That cannot happen to individuality. When Julius Caesar entered the Battle of Pharsalus (against Pompey), he didn't lose his identity; even if one could not identify him in the battle field, he is an individual. His identity does not depend on our ability to single him out in the middle of other persons. Quanta do not have identity and lack a principle of individuality, although they may be singled out (isolated) sometimes. 

Dieks made also a criticism to the interpretation given in \cite{kra11} according to which the trapped positron Priscilla could be substituted by any other positron of the universe and the measured results would be exactly the same. But it seems that this is precisely what happens to quantum objects in general! Of course Priscilla is \ita{that} positron in Dehmelt's trap, while it is there, but the argumentation in \cite{kra11} goes further, in supposing that during the night, when Dehmelt is at home, a malignant enemy destroys the experiment. But, moments later, the cordial caretaker, who saw Dehmelt making thousand of experiments of trapping positrons, until he decided to fix in one of them to call it `Priscilla' (any other could in principle do the job as well,  something that could not occur with individuals due to their differences), makes the experiment again just to please the Professor, trapping a positron (we can never come even close to establish that it is Priscilla again). Well, in the morning we shall have a positron in the trap, and Dehmelt, arriving at the lab, can say, as he does every day just for fun: `Good morning, Priscilla'. Well, neither Dehmelt nor Dieks could say that \ita{that} positron is not the Priscilla of the day before without knowing what has happened. It is completely different if one day, during the time Julius Caesar remained captured by pirates, early in the morning there arrives at the deck not Julius Caesar, but Pompey. That's the difference! Priscilla has not identity, but Julius Caesar does (so tells us our preferred metaphysics). 

\section{The classical theory of identity and its consequences}\label{sec4}

Roughly speaking, identity is a logical notion, while, let us recall, individuality is a metaphysical notion and individuation is an epistemological one. We have a \ita{prima facie} informal attempt at a characterisation of identity when we say that an object is identical just to itself and to nothing else. This \ita{metaphysical identity} (or \ita{numerical} identity) cannot be formalised in first-order languages, as mentioned above, and such an intuitive and redundant concept is not useful at all in dealing with foundational questions, yet sometimes it can be used for doing physics at a certain limit.\footnote{Concerning mathematics, it was mainly in the XIXth  century that the need for precision of concepts was seen  as necessary; as is well known, logic grew up with such a move. Perhaps it is time for the philosophy of physics to pay attention to the underlying logic of the physical theories.} So, we need to restrict ourselves to the theories we can develop regarding the concept of identity, and there is not one sole way to consider this concept.\footnote{We can recall Geach's relative identity, where we consider always $a$ being identical to $b$ relative to some sortal predicate $P$ (see \cite{noo15}), Hilbert and Bernays' definition of identity assumed by Quine, as a kind of indiscernibility relation by all the (finitely many) predicates of the language \cite{qui86}.} In other words, we need to make use of some logical system, and identity becomes relative to it, something (to our knowledge) almost never considered by philosophers of physics, who quite always work with the intuitive notion, thinking that it can be applied to everything. Thus, it is interesting to see what we have from the logical point of view if we adopt, as most physical theories do, classical logic as the underlying mathematical apparatus. This leads us to the CTI, the `classical theory of identity' to be described now, and to its consequences for the RV.

The central point in our argumentation concerning logic and mathematics is related to the fact that such apparatuses are used to express quantum mechanics; as we shall see, any physical theory is grounded on a logic and on a mathematics, usually taken to be that one which can be expressed in a theory of sets such as the ZFC system. 

Within ZFC, if we are going to represent the idea of a lack of identity, the best we can do is to `veil' the identity of some objects, say by confining them within an equivalence class or, in a most general situation, within a \ita{deformable} (non-rigid) structure. But this is a trick; we are just not `seeing' them as distinct, but they are individuals in the end.\footnote{A deformable structure has automorphisms other than the identity function. Two elements $a$ and $b$ of the domain of the structure are indiscernible (relative to the structure) if and only if there is an automorphism $h$ of the structure such that $h(a) = b$. It can be proven \cite[p.66]{jec03} that the standard universe of sets is rigid, meaning that its only automorphism is the identity function --- which entails that an object whatever represented in such a framework is indiscernible only from itself. For details, see \cite[chap.7]{frekra06}.} 

The defence we do of the RV is not to be taken as dogmatic. We are aware that there are alternatives, and nothing more than pragmatic criteria can be used to chose among these alternatives. But we are yet not convinced that the RV is not the better one to consider the very nature of the `particles' in quantum theories. 

The first is the underlying theory of identity, which we term the Classical Theory of Identity (CTI). If the binary symbol `=' is a primitive logical symbol of our language, the postulates of CTI are:

\begin{enumerate}
    \item (Reflexivity) $\forall x (x = x)$
    \item (Substitutivity) $x = y \to (\alpha(x) \to \alpha(y))$, where $\alpha($) is a formula where $x$ appears free and $\alpha(y)$ is got from the previous one by substituting $x$ for $y$ in some free occurrences of $y$ and $y$ is a variable distinct from $x$.
    \item (Extensionality-1) $\forall z (z \in x \lra z \in y) \to x=y$, in the case of `pure' sets (without ur-elements), or
    \item (Extensionality-2) $\forall z (z \in x \lra z \in y) \vee \forall z (x \in z \lra  \in z) \to x = y$ if the theory encompasses ur-elements. 
\end{enumerate}

The converses of last two follow from Substitutivity. Let us consider a universe of sets. As a first order theory, ZFC (supposed consistent) has infinitely many models of different cardinalities; it is not categorical. Let us fix in von Neumann's well-founded universe $\mathcal{V} = \langle V, \in \rangle$, where $\in$ is the membership relation and $V$ is the cumulative hierarchy of sets (either comprising ur-elements or not). Of course $\mathcal{V}$ is not a ZFC-set, that is, a set whose existence is obtained from the axioms of the theory; it is a proper class. But, in the metamathematics, it can be seen as a structure and the following theorem can be proven \cite[pp.66-7]{jec03}: \ita{in $\mathcal{V}$, there are no nontrivial automorphisms}. This means that the structure is rigid, and that the only object indiscernible from an element $a$ of the domain is $a$ itself. 

Really, given $a$, we need to acknowledge that the singleton $\{a\}$ can be formed (by the pair axiom); thus define the predicate $I_a(x) := x \in \{a\}$. Then it is easy to see that only $a$ obeys this monadic predicate, hence being distinct from any `other' object, that is, any object that doesn't obey $I_a$. 

Some philosophers prefer to say that quantum objects, although they cannot be \ita{absolutely discernible} (discernible by a monadic property), can be \ita{weakly discernible} (see \cite{mulsau08}, \cite{diever08}). CTI and the predicate $I_a$ show that this is a mistake. The language we use in developing quantum mechanics, even if it is not confined to predicates of the theory, ends up getting in the way of establishing identity and discernibility. 
 By `weakly indiscernible' we mean that objects $a$ and $b$ obey an irreflexive and symmetric relation (in the case of quantum objects, the two electrons of an Helium atom in its fundamental state for instance, consider the relation `to have the spin in the same direction of'), and by being \ita{only} weakly indiscernible, we mean that they cannot be distinguished absolutely. As shown in the previous paragraph, the monadic predicate $I_a$ distinguishes $a$ from any other object of the universe. So, in ZFC, there are not just weakly indiscernible things, and this applies to quanta once we ground our mathematics in such a set theory. 

\section{The money in a bank account analogy}\label{sec5}

In discussing the individuality of quantum objects, Schr\"odinger made a comparison of the metaphysical character of non-individuals with a case of an amount of money in a bank account \cite{sch98}. 
According to him, if we have, (say) \EUR 2 in a bank account, there is no sense in asking for \ita{our particular} euros; no single \EUR 2 coin is the one that we could say is our euro coin currently in the account. His conclusion is that the euros in a bank account are behaving as not individuals, \ita{at least for the purposes of attribution of a quantity of euros for a given client of the bank}. This would be so with quantum entities too: only the amount of them (and their species) would be relevant for the purposes of an adequate quantum description (that is, for a description of a state according to quantum mechanics). For quantum mechanical purposes, that is, all that matters is the kind of the entities involved and their quantity, not which of them are being used.

In analysing this problem, Dennis Dieks says that 

\begin{quote}
    more than one money units in a bank account is the standard example of the absence of \ita{individuality}; it is a case in which only the account itself, with the total amount of money in it, can be treated as possessing \ita{individuality}. (our emphasis) \cite[p.53]{die14}
\end{quote}

We have emphasised the word `individuality' to call the attention, again, to the (\ita{pace}, Dennis) confusion among identity, individuality and individuation. If I have just one euro in my account, then of course there needs still be no individuality here in the form of a principle of individuality, but individuation can be considered, since \ita{we} can say that \ita{that} euro in \ita{my} account is mine. But what lacks here is identity. Since any transference between two accounts can be done, we cannot even more identify `my' euro if it is deposited in another account with more euros there. 

Situations like these ones, where quantity matters, but individuality and identity do not, are precisely what the theory of quasi-sets  captures. But, before seeing how this can be done, let us remark that if the analogy is to be put within a standard language such as the language of ZFC, we can ask: is the collection of the euros in our account a \ita{set}? Remember that, by hypothesis, we are developing our physical theory as having classical logic (meaning ZFC) as its underlying logic. So, everything is either a set or an ur-element (if there are any) in this context. Whatever the answer, the money in a bank account ends up having a kind of individuality, as related to the classical theory of identity, which we have briefly discussed before. 

Even if that is clear enough, some, such as Dieks, suggest that to resort to ZFC in this discussion would be ``an extraordinary and highly artificial measure'' \cite{die20}. He goes on and says that bankers and clients need only classical logic and set theory, and that even withing such a framework we can reason as if there are discrete sums in the account and this does not imply that there are individual euros in it. We don't agree. If bankers and clients don't want to identify their particular euros while they are in their accounts, ZFC is just what they \ita{shouldn't} use!

Really, the hypothesis that there are no individual euros in the account presupposes that one is ignoring the mandatory postulates of ZFC. Of course bankers and clients don't use formal ZFC, and we cannot sustain this view, for they certainly do not even think about shifting to a formal system of set theory to deal with their money. They reason intuitively \ita{as if} the particular euros could not be identified and had no specific individuality when they make financial operations on them. But whenever one tries to formalise this operation, one will surely find that we need to consider the euros as forming a set of euros, and then their identity enters by the backdoor. Dieks still refers to the fact that sometimes certain amounts of money can acquire identity, for instance when we have just \EUR 1 in our account. This case is completely similar to the situation involving trapped quanta. You can call `Priscilla' your euro, and the same remarks made above apply also here. Specifically with the money case, what we can do (say, to buy a caramel) with one euro, we can do with any other one, taken from a different account named by another person (which, by hypothesis, does have identity). They will acquire identity only when you go to the bank and get it in the cash; then you can say `This one is \ita{my} euro', and it is \ita{different} from any other one in the world. But now you are speaking of a particular piece of paper, and not of the euro as it act in finances. No, even one euro is not an individual in the sense we use this word when only financial operations are involved, and few people who discuss this subject says what they understand by an individual or by `having identity', Dieks included. 

As the previous discussion makes it clear, the money in a bank account is an analogy. From the point of view of the uses of money and for financial purposes, amounts of money are not individuals. They may be permuted without any loss of their functions. There is no specific financial transaction that can be done with one coin of \EUR 2 that cannot be done also with another coin. So, from this point of view, they have no individuality. Certainly, one can provide identification for money, by describing amounts of money relatively to well-identified objects, such as `the \EUR 2 coin in my left pocket', or even `the coins in my safe'. This provides for identification, but not for individuality, if are considering only the financial operations. However, that is not the only perspective from which coins and money in general can be seen. As macroscopic objects, they may be treated as physical systems. That guarantees, in particular, that each single coin has a specific position in space, so that it gets individuality, from that point of view, as a physical object. Certainly, in ordinary circumstances we do not keep separating contexts, such as those involving financial operations, and those involving physical objects. But again, the point of the analogy was that, when we restrict ourselves to a given kind of description of given objects, they may be seen as non-individuals. 

What happens in quantum mechanics is that, differently from money in a bank account, where one also has (classical) physical description available, our access to quantum entities is granted exclusively by quantum mechanics. One just cannot see quantum particles as also classical objects, in the sense that such objects could also be described by classical physics; the best we can have is some occasions of identification, but that does not grant individuality, as we have already discussed. Being quantum objects, and granting that the quantum description does not allow for any kind of natural individuality principle (no Leibnizian principle, given the permutation symmetry, and assuming that Transcendental Individuality is a philosophical posit, not a quantum mechanical one), the kind of operation of granting individuality through other means, such as happens in the case of the money, is lost. From a quantum mechanical perspective, it is as if we only had the financial description of the coins. 

Quasi-set theory does the job quite nicely by considering non-individuals within the Received View. Let us see how.

\section{Quasi-sets, an intuitive view}\label{sec6}

Let us provide a general view on quasi-sets (qsets); details can be seen in \cite[chap.7]{frekra06}, \cite{frekra10}. Quasi-sets are collections of objects some of which may lack standard identity conditions or, better saying, the CTI doesn't apply to them. Other objects may obey CTI, and are called `classical objects' in the theory. Quasi-sets with only classical elements are \ita{sets}, and in this case the theory reduces to a standard set theory. There are two kinds of atoms (ur-elements), but of course we can enlarge the theory to admit more; the atoms are divided in two kinds, $m$-atoms, which play the role of quantum entities, and the $M$-atoms, which play a role similar to the ur-elements of ZFU, the Zermelo-Fraenkel set theory with ur-elements. The logical vocabulary is standard of classical first-order logic, but no primitive identity sign is assumed. Instead of identity, the language has as primitive a binary predicate `$\equiv$' for \ita{indistinguishability}, or \ita{indiscernibility}, which has the properties of an equivalence relation (but it does not obey full substitutivity). A concept of \ita{extensional identity} (or just `identity' for short) is defined: $x=y$ holds for $M$-atoms belonging to the same qsets or to qsets having the same elements. But notice that since the standard identity doesn't hold for $m$-atoms, we cannot suggest an exercise of verifying if a certain $m$-atom belongs to a certain qset. This is as if you would like to know if a certain quark (suppose you can think of it, say the only down quark in an Hydrogen atom) is or not \ita{that} quark of one of the Hydrogen atoms of a water molecule.\footnote{Notice the difficulty in trying to give such an example; of which specific water molecule we are speaking about? Of which Hydrogen atom? This difficulty reinforces the idea that these entities do not have a well defined identity.}

Although we cannot assert \ita{which} $m$-atoms are the specific elements of a qset, we may reason as if there are some, and then the notion of cardinal applies also here. A Hydrogen atom has one proton and one electron, but no one will ask \ita{which ones}. We can think of such an atom as a qset whose cardinal is 2 and whose elements form sub-qsets of cardinality one. The theory proves that, given a certain $m$-atom, there exists a qset with cardinal one whose only element is indistinguishable from the given element. But we cannot prove that the element is \ita{that} given element for the proof relies on identity. But this enables us to speak, say, of the only electron in the outer shell of a Sodium atom without the consequences of being able to identify it as if we were within a standard framework. Really, suppose we are in ZFC. Then we should regard the electrons of a Sodium atom as a set  with 11 elements, and all of them would be discernible (due to CTI), in particular the electron in the outer shell. In the quasi-set theory, we can reason as we do with the two electrons of an Helium atom (a most common example) in the fundamental state: we know that there are two and that they differ by their spins in a given direction, but we cannot tell which is which. The same with the Sodium atom: there is just one electron in the $3s$ shell, but nothing tells us which electron is it. 

Quasi-set theory describes quite nicely the statistics. As already shown in \cite{sankravol99}, \cite[chap.7]{frekra06}, in considering indiscernible $m$-atoms we arrive `directly' to quantum statistics without the (standard) necessary \ita{ad hoc} assumption that the quanta of the same kind are indiscernible; remember that we have started with such an hypothesis as a metaphysical assumption. So, the Indistinguishability Postulate is not necessary.\footnote{Perhaps something more should be added here. There is a difference, a fundamental one according to our concerns, in which we start with some metaphysical thesis we either believe or wish to investigate (our case) on the one hand, and to make an \ita{ad hoc} hypothesis after the ship having left the port (the introduction of the Indistinguishability Postulate), on the other.} 

An important remark is in order. One of the most subtle criticism to the theory is that, in its own formulation, identity is assumed \cite{bue14}. Our answer is as follows (details in \cite{kraare19}). 
Some paraconsistent logics violate the Principle of Non-Contradiction, accepting that $\alpha \wedge \neg \alpha$ (for some $\alpha$) can be true sometimes (in these logics, the explosion principle $\alpha \wedge \neg\alpha \to \beta$ does not hold). But in describing such logics, we do not violate such a principle. Really, we never think that something can simultaneously be a formula and not be a formula, so as that the symbol `$\neg$' stands for negation but also does not. Classical logic holds in the metalanguage, and so it does in the case of quasi-sets. Identity holds when one is developing the formal system of quasi-set theory. The theory used to construct and develop a formal system needs not share the same principles the theory in the object level requires. For the purposes of developing formal systems, a constructive logic is enough. That does not mean that the object level system needs to be constructive. Something similar holds for systems restricting identity; although at the metalevel we need identity, the target to be investigated with the object system is circumstances where the laws of identity do not apply, and this is a different domain.

A (finite)\footnote{Finite quasi-sets apparently are enough for the relevant applications we have in mind.} quasi-set (qset) can be seen as a tuple 

\begin{eqnarray}\label{eqn_q}
q = \langle k_1, \ldots, k_n ; \lambda_1, \ldots, \lambda_n \rangle,
\end{eqnarray}

\noindent where the $k_i$ indicate the `kinds' of the elements of the qset, while $\lambda_i$ the cardinality of the kind $k_i$; the theory shows how to associate cardinals to quasi-sets. For instance, the Helium-3 isotope, for certain purposes, can be written as 
\begin{eqnarray}
\mathrm{He}_3 = \langle p,n,e ; 2,1,2\rangle,
\end{eqnarray}

\noindent with $p$, $n$ and $e$ standing for `protons', `neutrons' and `electrons' respectively, and the cardinals indicating the quantities of each one. This intuitively means that we have two protons, two electrons, and one neutron, and this takes not only these entities as indiscernible but two He$_3$ atoms as well. The case with just one euro is similar:
\begin{eqnarray}
E = \langle \mathrm{euro},1\rangle
\end{eqnarray}

\noindent says that we have just one euro, without identifying it. A qset with such characteristics is called \ita{a strong singleton}, expressing one unity of a kind, without identification of the element (to see how this is possible, please see \cite[p.]{frekra06}).\footnote{See Section (\ref{sect_msets}) below a comment about multisets.} 

This kind of apparatus may provide for an interesting alternative to situations where ZFC is used. We do not mean by that that ZFC should be substituted or that it is wrong; rather, our claim is that for a specific purpose of dealing with quanta, perhaps, when the non-standard form of metaphysics we are advancing here is taken into account, then, an alternative system is more appropriate. For classical mathematics, however, ZFC is still more appropriate, and, as we have mentioned, qsets recover classical set theory when only classical objects are involved. 
The kinds of quanta, that is, the differences among quanta of the $k_i$ species are given by the physical theory and it is not a logical task to provide such a thing. Dieks and Versteegh say that ``there are no euros independently of account numbers, and possessing definite values'' \cite{diever08}, also mentioned in \cite{die20}. We think differently: without euros (or other currency), there cannot exist euros in accounts, and even the concept of a bank account would lose its meaning. Things come first, even if only in a metaphysical sense.\footnote{By the way, contrary to Quine, who claimed that there are no entities without identity quasi-set theory shows that it can be made sense of the idea that there can exist entities without identity. On this matter, we prefer to follow Barcan-Marcus, to whom, on the contrary, there is no identity without entity \cite[p.172]{frekra06}.} 

One of the most important effects of the development of a quasi-set theory, we believe, concerns a discussion on methodology of metaphysics of quantum mechanics. Given a claim that some entities may exist without individuality, and that this also means not having identity, the theory allows us to formulate it in precise terms, and to recover important results of quantum mechanics (see \cite{kraare16}, where it is shown that for technical purposes, a version of the theory is recovered in quasi-set theory, using the specific resources of the theory). Most objections to the RV concern the informal use of concepts like identity and cardinality. It is typically claimed (as by Dorato and Morganti \cite{dormor13}, and by Jantzen \cite{jan11}, for instance) that the concept of cardinality, the very idea of having a specific number of entities in a given circumstance, implies identity, even if only in a thin sense. Dorato and Morganti \cite[p.606]{dormor13}

\begin{quote}
one could maintain that the `presence' of $n$ particles at the formal level has a direct ontological counterpart, so that it can be concluded that those particles are $n$ individuals \ita{independently of their qualities}. After all, if the fact that a given physical system is composed of a certain number of particles in a purely formal sense is fundamental for even starting to show that the entities composing the system are discernible, it seems perfectly legitimate to regard quantum particles as (`lowdegree') individuals by moving from a purely formal to a non-formal reading of countability, without caring about qualitative properties and (in)discernibility.
\end{quote}

However, as we are mentioning, the informal concepts used in informal language, are just too imprecise to allow us to judge on such issues. For the sake of clarity, is always useful, for philosophical purposes, at least, to present a kind of regimentation of the concepts involved, which would bring to light the presuppositions involved in our intended use of the concepts and which are the consequences of such use of these principles. In this sense, we are not claiming that, once formulated classically, the idea of cardinality, counting, and identity, are intertwined, justifying the claim that counting requires and implies identity for the counted entities. However, given the imprecise nature of such concepts in natural language, they may also be formulated in distinct terms, the ones of quasi-set theory, where the kind of relation between counting, identity and cardinality is broken. In this setting, the claim holds that one may have cardinality without identity (see also \cite{arekra14}). So, the dispute cannot concern a preferred use of concepts in natural language. As a matter of formulation of metaphysical theories, the theory of non-individuals cannot be blamed to be faulty on those scores. As we have been advancing in this paper, it can avoid the misunderstandings if the proper conceptual distinctions are made, and quasi-set theory is a formal tool helping us on that score. The problem on the dispute concerning a metaphysics of individuals against non-individuals then, turns not on the specific principles assumed by each view and on their coherency and consistency, but rather on a dispute on the most appropriate metaphysical theory to be chosen.

\subsection{A note on multisets and fuzzy sets}\label{sect_msets}

The consideration of a quasi-set as something as given by equation (\ref{eqn_q}) may bring to some reader the notion of \ita{multisets} \cite{bli89}. At a first glance, it may seem that the idea is quite similar to that of quasi-sets, but this is not so, for there are important differences. A multiset is a collection of objects that enables the repetition of the elements. So, $\{1,1,2,3,3\}$ is a multiset (mset for short) with cardinal 5, and it is not identical to its `support set' $\{1,2,3\}$ which has cardinal 3 as in standard ZFC. So, a mset can also be represented by something like (\ref{eqn_q}), but in this case all elements of a same kind are \ita{the same object}, something that does not happen in the case of quasi-sets.\footnote{But notice that quasi-set theory does not say the opposite, namely, that the elements of a same kind are all different one each other. The theory doesn't enter in this question, for it depends on the physical theory in consideration. In this sense, a mset may be viewed as a particular case of a qset when identity holds for every entity and repetitions are allowed.}

So, we think that the claim that msets can cope with quantum objects is not correct, for in the case of the isotope He$_3$, no physicist will say that the two electrons or the two protons are \ita{the same electrons or protons} (`identical' in the philosophical jargon). We have analysed this case in \cite{kra91}. 

As for fuzzy sets, things are not deeply different. Roughly speaking, in a fuzzy set we have other alternatives for an element to either belong or not to a fuzzy set \cite{zah65}. Given a standard (`Cantorian') set $A$ and an object $a$ whatever, we have that either the object does belong to the set or it does not belong to it (excluded middle law). This can be seen from the point of view of the characteristic function of the set: $\chi_A(x) = 1$ if $x \in A$, and $\chi_A(x) = 0$ if $x \notin A$. So, the image of the function is the set $\{0,1\}$. Let us write $x \in^0 A$ to say that $x$ does not belong to $A$ and $x \in^1 A$ for $x$ belongs to $A$.
In a fuzzy set, the image is the interval $[0,1]$, so we can have $x \in^k A$ for any $k \in [0,1]$. Of course the extremes express the Cantorian case, so standard sets are particular cases of fuzzy sets. But we may have also $x \in^{0.3} A$ or $x \in^{0.7} A$ and interpret them as `$x$ belonging \ita{less} to $A$' and `$x$ belonging \ita{more} to $A$' respectively. 

This is of course a very nice idea, as the literature has shown. But concerning quantum objects, we have again a flaw in thinking that an electron, say, may belong \ita{less} to an atom than `another' electron. Really, the elements of a fuzzy set still obey the classical theory of identity\index{CTI} and hence are \ita{individuals}, presenting identity in the sense we have described before. So, what we have here is a kind of epistemological fault, for we have no means (which despite of this may exist) to know whether a certain element does belong or not to a fuzzy set. The quantum case is of course different. 

According to the main interpretations, that is, those which accept Bell's theorem which says that no local hidden variable theory can agree with quantum mechanics, there is no way of accepting that we can supply the theory in order to surpass such an epistemological fault and even so keep with the classical \ita{realistic} account. As J. -M. and and F. Balibar have suggested \cite[p.69]{levbal96}, perhaps we are in the presence of something radically different from the objects we are accustomed with; they suggested to call them (perhaps inspired in Mario Bunge) \ita{quantons}. 

\section{A final remark on metaphysical choice}\label{sec7}

What we have reached, then, with our discussions on quasi-set theory, is that one may have plausible formulations of the concepts involved in the articulation of the RV, without them presupposing a relation with identity as most critics suggest. What is at stake is that one may formulate the cluster of concepts involved in articulating the RV and its rivals in distinct ways, some of which do involve the relation between identity and cardinality (to stick to one example), and some of them do not. Also, some approaches to metaphysics do conflate on purpose identity and individuality, by explaining individuality in qualitative terms (the Leibnizian approach), and reducing identity to indiscernibility by the set of predicates of a language. Other approaches, however, avoid such conflation. For the purposes of the RV, at least as typically formulated, keeping the concepts apart is the most interesting option. 

This leads us to a step back in the discussions that have been discussed so far. The issue is not whether the theory of quasi-sets is plausible or adequate, or whether one can use a particular concept of individuality or cardinality. Certainly, quantum mechanics will work for most of them (not simple versions of the Leibnizian view; see French and Krause \cite[chap.4]{frekra06}). The problem concerns metaphysical theory choice: which view to adopt. We shall make no remarks on whether metaphysics aims at truth about reality or just on developing beautiful theories (see Benovsky \cite{ben16} for further discussion). Quantum mechanics cannot tell us which metaphysical theory to prefer, the metaphysics of individuality gets underdetermined by the theory (see also French and Krause \cite[chap.4]{frekra06}). Our purpose here was to argue that some pre-conceptions operating on the use of concepts such as cardinality, individuality and identity cannot rule a metaphysics of non-individuality also. Once properly understood, such concepts may still play the roles they are expected to play, and not get in the way of the development of a metaphysics of non-individuality in quantum mechanics. 

One could now attempt to go one step further. Given that quantum mechanics does not allow for a straightforward Leibnizian theory of individuality, and also, given that Transcendental Individuality seems to be a gratuitous addition over the physics, provided merely in order to satisfy some philosophers' needs for a principle of individuality, wouldn't it be better just to keep without any principle of individuality at all? This, we suggest, would be closer to quantum mechanics, and also could be seen as a further lesson of the theory. Quantum objects are not local (Bell's theorem), they are not complete (in the sense of having or else lacking every property, due to Kochen-Specker theorem). Why not say that they also have no individuality, due to permutation symmetry? Methodologically, the suggestion has many attractions for a metaphysical naturalist, but, interesting as it is, we shall develop it elsewhere.

\end{document}